\documentclass[twocolumn,showpacs,superscriptaddress,aps,floatfix]{revtex4}

\usepackage{graphicx}
\usepackage{rotating}
\usepackage{amsmath}
\usepackage{color}
\usepackage{graphicx}
\usepackage{makeidx}
\usepackage[sort&compress]{natbib}

\newcommand{\be}{\begin{equation}}
\newcommand{\ee}{\end{equation}}
\newcommand{\bea}{\begin{eqnarray}}
\newcommand{\eea}{\end{eqnarray}}

\newcommand{\etsf}{European Theoretical Spectroscopy Facility (ETSF),Centro Joxe Mari Korta, Avenida de Tolosa, 72, , 20018 San Sebasti\'{a}n, Spain }
\newcommand{\sanseb}{Departamento de F\'isica de Materiales,
Facultad de Qu\'imicas, 
Unidad de F\'isica de Materiales Centro Mixto, 
Donostia International Physics Center, Universidad del Pa\'is Vasco, San
Sebasti\'{a}n, Spain}
\newcommand{\austria}{Faculty of Physics, Vienna University,
Strudlhofgasse 4, 1090 Wien, Austria}
\newcommand{\dresden}{IFW Dresden, P.O. Box 270116, D- 01171 Dresden,
Germany}

\begin{document}
\title{\emph{Ab-initio} band structure of doped graphene}
\author{C. Attaccalite}
\affiliation{\sanseb}
\affiliation{\etsf}

\author{A. Gr\"ueneis}
\affiliation{\austria}
\affiliation{\dresden}
\author{T. Pichler}
\affiliation{\austria}
\affiliation{\dresden}
 
\author{A. Rubio}
\affiliation{\sanseb}
\affiliation{\etsf}

\begin{abstract}
We present an  \emph{ab-initio} study of the graphene quasi-particle band structure as function of the doping in $GW$ approximation.  We show that the LDA Fermi velocity is substantially renormalized and this renormalization rapidly decreases as function of the doping. 
We found, in agreement with previous papers, that close to the Dirac point the linear dispersion of the bands is broken but this behavior disappears with a small doping. We discuss our results in the light of recent experiments on graphene and intercalate graphite.
\end{abstract}           

\maketitle

\section{Introduction}
Since its discovery \cite{novoselov1,novoselov2,zhang} graphene raised a great interest in the scientific community for the possibility to use it as new material with tunable properties \cite{nano} in electronic devices and to probe open quantum mechanics questions.   Electron-electron interaction in graphene, and related compounds, are expected to play an important role due to its low dimensionality. Lately the quasi-particle dynamics of graphene and graphite has been addressed by high-resolution angle resolved photoemission spectroscopy ARPES \cite{bostwick,alexander,alexander2} experiments. It has been found that conical bands are distorted due to many-body interactions, which renormalizes the Fermi velocity $v_F$. The self-energy corrections to the bare band structure are crucial for determine transport,  optical properties and other properties of graphene based materials.\\ 
Moreover recent experiments on doped graphene \cite{rotenberg} report variation of the Fermi velocity and  Fermi surface topology induced by doping. Minor changes are definitely due to the small variation of the lattice constant \cite{pietronero} with the doping but mainly it is expected that larger differences are due to correlation effects in doped graphene \cite{sarma,polini}. 
In this paper we study the quasi-particle band structure of extrinsic graphene (doped graphene) as function of the doping, gated or electrostatic. Correlation effects are evaluated using \emph{ab-initio} many-body self-energy calculated in $GW$ approximation, where the polarization needed to evaluate $W$ is calculated within the Random Phase Approximation.  The validity of the RPA in the calculation of graphene self-energy has been recently questioned \cite{mishchenko} but it has successively  shown to be a valid and controlled approximation in doped graphene for any sufficient small doping  \cite{sarma}. Therefore in this paper we always introduce a small broadening in the Fermi distribution of the electrons even in the undoped case in order to guarantee the validity of our approximations and be closer to the experimental results where a finite doping is always present, due to impurities or substrate interaction.
\section{Methods}
In order to simulate isolated graphene we used a slab-geometry, i.e., bulk geometry with large distance between the layers, and an elementary $2D$ graphene cell with 2 atoms with a lattice constant $a=2.46$ Angstrom. DFT-LDA calculations were performed with the plane wave-code ABINIT \cite{ABINIT}, using a cutoff of $60Ry$ for the wave-function, a  $42x42x1$ Monkhosrt-Pack grid for the density and Troullier Martins pseudo-potentials \cite{pseudo} for the carbon atom. 
 The quasi-particle band structure were successively obtained from the self energy $\Sigma(E)$ evaluated in the $GW$ approximation \cite{rubio,gunnarsson}.  $GW$ calculations were performed with the code $Yambo$ \cite{yambo}  starting from DFT-LDA wave-functions. 
      Because we are not interested in the total energy but only in spectral properties, we didn't perform a self-consisted calculation but stop this approximation to the first order $\Sigma = G_0 W_0$ \cite{gwfull}. The quasi-particle band structure is then obtained as:
\be
E_{nk} = \epsilon_{nk} + Z_{nk} Re\Delta \Sigma_{nk}(\epsilon_{nk})
\label{per}
\ee
where $\Delta \Sigma = \Sigma-V^{xc}$, $V^{xc}$ is the LDA exchange-correlation functional, $\epsilon_{nk}$ are the Kohn-Sham eigenvalues and Z is the renormalization factor $Z_{nk} = [1-\partial Re \Delta \Sigma / \partial w]^{-1}$.    The screened electron-electron interaction $W$ have been calculated within random phase approximation (RPA) in term of the dielectric constants $\epsilon_{G,G'}(q,\omega)$ using a plasmon-pole model \cite{ppmodel}.  Bands up to $60eV$ in energy have been employed both for $W$ and for the expansion of the Green Function $G$, that correspond roughly to $50$ bands in the case of $20$ a.u. distance  between the graphene sheets. Convergence in the size of the matrix $\epsilon_{G,G'}(q,\omega)$ has been carefully checked. In all the $GW$ calculations a small  Fermi-Dirac broadening $0,002 Ry$ was used. The dependence of the $GW$ results from the k-point sampling and the distance between the sheets is discussed in the Appendix A.

\section{Quasi Particle Band Structure of Graphene}

\begin{figure}[ht]
\begin{center}
\includegraphics[width=0.33\textwidth,angle=-90]{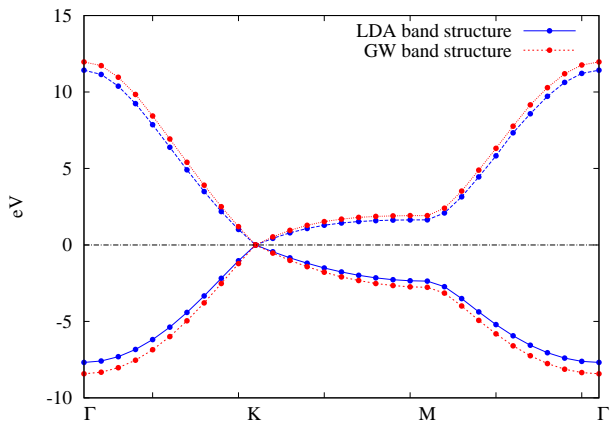}
\caption{Band Structure of Graphene in LDA and $GW$ approximation, using $20$ a.u. slab geometry and $36x36x1$ kpoint sampling. \label{gw_graphene}
}
\end{center}
\end{figure}

We report the quasi-particle band structure of graphene in fig. \ref{gw_graphene} for the $\pi$ and $\pi^*$ bands along the symmetry lines \boldmath$\Gamma KM\Gamma $  \unboldmath in LDA and $GW$ approximation.  It's clear that the bandwidth is increased  in $GW$ and consequently the Fermi velocity $v_F$ is renormalized. We report $v_F$ along the direction \boldmath$K\Gamma$ and $KM$ \unboldmath in the table I, extrapolated for an isolated slab of graphene (see Appendix A). 

\begin{figure}[ht]
\begin{center}
\includegraphics[width=0.33\textwidth,angle=-90]{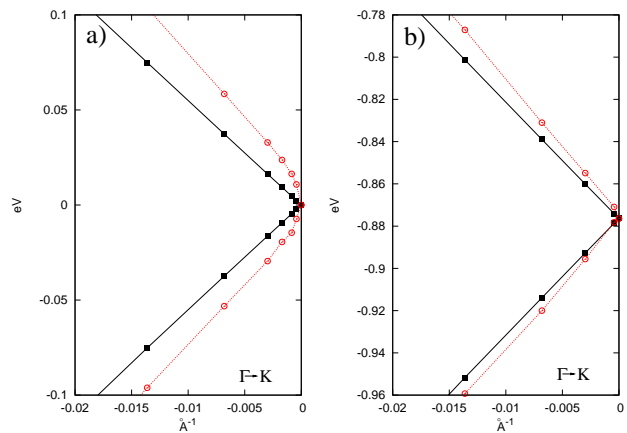}
\caption{Band Structure of Graphene in LDA and $GW$ approximation close to the \boldmath$K$\unboldmath point \label{closek}. (a) Undoped graphene, (b) doped graphene with $0.05$ 
electrons in excess in the primitive cell.}
\end{center}
\end{figure}

Very close to \boldmath$K$\unboldmath~we found a non-linear behavior of the $\pi$ bands similar to the one obtained by Trevisanutto et al. ref. \cite{olevano}, see fig. \ref{closek} panel $a)$. However we notice  that the slope of the bands we obtained close to \boldmath$K$\unboldmath~is slightly different and this can be due to the plasmon-pole approximation employed in the calculations, see also ref. \cite{olevano}.
We then studied the anisotropy of the bands close to $K$ in $GW$ and LDA. We evaluate  the Fermi velocity in different directions and found that the  $GW$ corrections increase the anisotropy respect to the LDA in undoped graphene. The results for $v_F$ in different directions extrapolated for isolated graphene are reported in table  \ref{table1}.\\

\begin{table}[t]%
\begin{tabular}
[c]{|l|c|c|}\hline
 & LDA & $GW$  \\ \hline
\boldmath $KM$  & $0.802(1)$ & $ 1.00(5)$ \\
\hline
\boldmath$K\Gamma$         & $0.860(2)$ & $ 1.15(5)$ \\
\hline
\end{tabular}
\caption{Fermi velocity of graphene extrapolated for infinite distance between the slabs along the direction \boldmath$K\Gamma$ and \boldmath$KM$.}%
\label{table1}%
\end{table}

\section{The effect of doping}
\begin{figure}[ht]
\begin{center}
\includegraphics[width=0.33\textwidth,angle=-90]{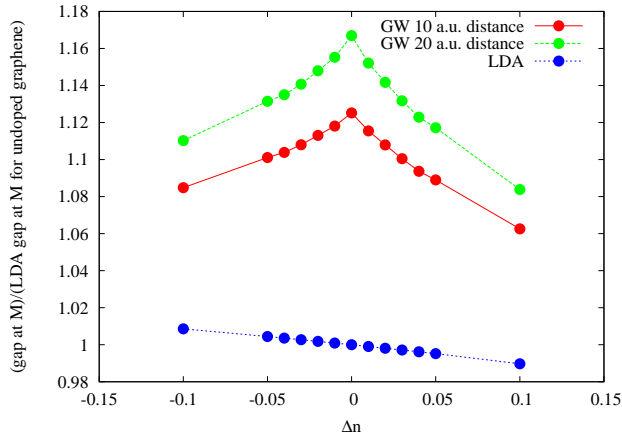}
\caption{Band gap of graphene at the \boldmath$M $ \unboldmath point in $GW$ and LDA as function of the doping \label{gw_m_doping}, divided for the LDA gap at zero doping.  We report the calculations for two different distances of the graphene slab.}
\end{center}
\end{figure}

\begin{figure}[ht]
\begin{center}
\includegraphics[width=0.33\textwidth,angle=-90]{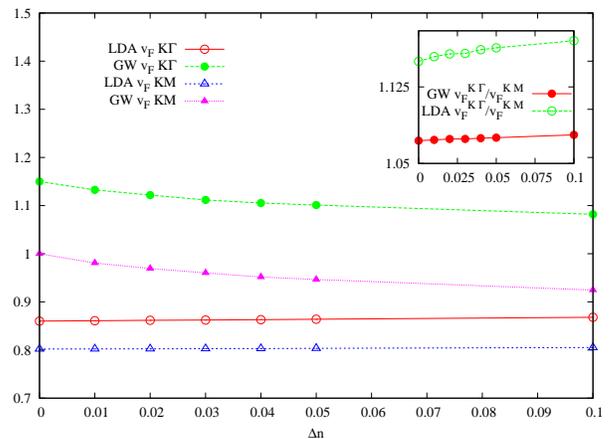}
\caption{Fermi velocity $v_F$ as function of the doping \label{vfermi}, in the inset we show the ratio of the Fermi velocity along the line \boldmath$K\Gamma$\unboldmath and \boldmath$KM$\unboldmath}
\end{center}
\end{figure}

Doped graphene is simulated changing the number of electrons in the unit cell and then compensating the negative charge with an uniform positive background. When we start to dope graphene the system evolves from a semi-mental to a real metal, and the effective interaction felt by the electrons starts to be weaker because of the stronger screening of the Coulomb potential. This causes a smaller renormalization of the band structure and a consequent decrease of the Fermi velocity.  In fig.  \ref{gw_m_doping} we report the gap at the M point of graphene versus the electrostatic doping, that is represented by $\Delta n$ the excess of electron in the unit cell, divided for the LDA gap at zero doping. As one can see while in LDA the gap slightly varies  with the doping, on the other hand the $GW$ renormalization is very strong for intrinsic graphene and decreases rapidly as function of the doping.  \\ 
Then we analyzed the band structure close to the Dirac point and we found that the kink rapidly disappears with doping. We report in the fig \ref{closek}, the two extreme cases the intrinsic graphene and the highly doped graphene where the kink disappears. This finding is in agreement with the interpretation of S.Y. Zhou et al. \cite{zhou} of the gap opening in graphene  due to the substrate \cite{giovannetti} and not to manybody effects. 
In fact even a small doping as the one present in the experiments of S.Y. Zhou et al. makes the kink too small to affect the ARPES results.
Notice however that using a pertubartive solution for the Dyson equation, see Eq. \ref{per}, we are not able to describe the plasmaron solution found by ref. \cite{polini,hwang}, so we cannot exclude many-body effects in the ARPES results. Moreover the screeening is evaluated  only in plasmon-pole approximation so  effects due to presence of additional poles in $\epsilon^{-1}$ are not considered (see ref. \cite{olevano} for a discussion).
\\
Then we studied the change in topology of the quasi-particle band structure with the doping, in order to clarify recent experiments on doped graphene and on graphite that found a large trigonal distortion in the band structure \cite{rotenberg}. In fig \ref{vfermi} we report the $v_F$ along the direction \boldmath$KM$\unboldmath and \boldmath$K\Gamma$\unboldmath as function of the doping in LDA and $GW$. As expected the $GW$ results slowly go closer and closer to the LDA ones for increasing doping.  Notice however that the change of the $v_F$ in the two directions as function of the doping is different. In the inset we report the ratio of the Fermi velocity along the line \boldmath$KM$\unboldmath and \boldmath$K\Gamma$\unboldmath. This value can be used as  measure of the Dirac cone anisotropy. We found that the anisotropy is larger in $GW$ and increases with the doping. A similar result was found also by Roldan et al. \cite{guinea} using tight-binding model, even if the increasing of the anisotropy they found is larger than ours. Finally we want to comment that in order to understand the size of trigonal wrapping found by ref.  \cite{rotenberg} it is necessary to consider the hybridization of carbon $\pi$ bands with  the dopant ones, that causes a distortion of the band structure close to the Fermi energy, see for instance ref.\cite{mauri}.   

\section{Conclusion}
We have studied the quasi-particle band structure of graphene as function of the doping. The bands dispersion is obtained by the real part of the self-energy calculated in $G_0W_0$ approximation, that is generally considered accurate for a broad range of materials, including graphene.  We found a strong renormalization for intrinsic graphene, that rapidly decrease as function of the doping.  This result can be used as new reference for future experiments on doped graphene and intercalate graphite because it provides for the first time an estimation of the Fermi velocity change with the doping. Close to the Dirac point we found the presence of a kink in the graphene band structure, but it rapidly disappears with doping, at least in $G_0 W_0$ approximation using a plasmon-pole model for $W_0$. This finding is in agreement with the experimental results and their interpretation by Zhou et al. \cite{zhou}. Finally we report the slope of the Fermi velocity in the directions \boldmath$KM$ and $K\Gamma$\unboldmath as function of the doping, and found that the trigonal wrapping is enhanced by the $GW$ corrections and it increases with the doping.

\section{Acknowledgment}    
A.R. and C.A. are supported in part by Spanish MEC (FIS2007-65702-C02-01), Grupos Consolidados UPV/EHU of the Basque Country Government (IT-319 -07) European Community e-I3 ETSF and SANES (NMP4-CT-2006-017310) projects. C. A. thanks Marco Polini, Valerio Olevano and Ludger Wirtz for useful discussions.

\appendix
\section{Convergence of the calculations}
\begin{figure}[ht]
\begin{center}
\includegraphics[width=0.33\textwidth,angle=-90]{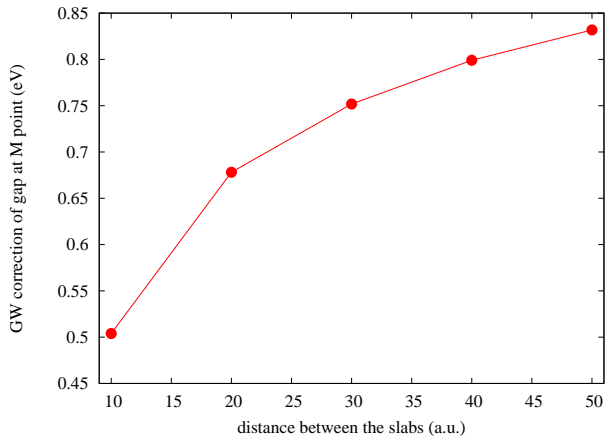}
\caption{Converge of $GW$ correction to the gap at \boldmath$M $ \unboldmath as function of the sheets distance. \label{m_dist}}
\end{center}
\end{figure}

In a typical $GW$ calculation different convergence parameters have to be carefully considered, here we want stress the relevance of the convergence in the k-point grid and the distance between the graphene sheets.  The major part of the $GW$ calculations are performed using a supercell approach, where in order to simulate an isolated system a large distance $d$ between the slab is used. It is well known that the $GW$ corrections for a $2D$ systems are larger because the screening of Coulomb potential is weaker making the electron-electron interaction stronger, but in a slab geometry this is true only in the limit of infinite distance between the slabs. Here we report the change in the $GW$ corrections at \boldmath$M$\unboldmath point as function of the distance fig. \ref{m_dist}. However notice that in the experiments the screening by the substrate will never allow for the observation of a completely single sheet, so the direct comparison with experiments remains more difficult.
As one can see from the fig. \ref{m_dist} in order to converge the calculations with an accuracy of $0.05$ eV a distance larger than $50$ a.u. is necessary, however in this paper we performed all calculations using a distance of $20$ a.u. mainly for two reasons: first the behavior of  the $GW$ corrections with doping remains unaltered; second calculations with distances equal or larger than $50$ a.u. are computationally unfeasible using plane-waves. Anyway even if we cannot provide the full band structure with large distance we used a fix grid of k-points to extrapolate the value of the Fermi velocity  for different sheet distances to the infinite one $v_F^{\infty}$. In table \ref{table1} we report the final result that was obtained fitting the $v_F$ with $v_F (d)=  v_F^{\infty} + A/d$, where $d$ is the distance between the sheets and $A$ and $v_F^{\infty}$ are fitting parameters.

\begin{figure}[ht]
\begin{center}
\includegraphics[width=0.33\textwidth,angle=-90]{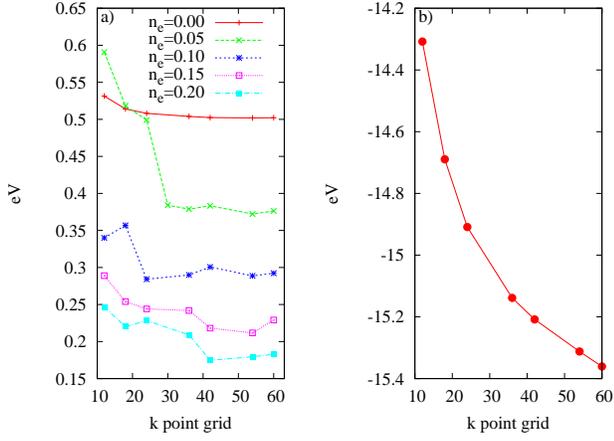}
\caption{$a)$ Converge test of the $GW$ correction to the gap at \boldmath$M$\unboldmath point in graphene as function of the k-point grid plotted for different number of electrons. In x label we report the size of the grid, namely $n$ stays for a grid $n \times  n \times 1 $. All results are converged within 0.01 eV. \label{conv_k}. $b)$ convergence of the $\Sigma^X$ for the $\pi$ band at the \boldmath$M $ \unboldmath point.
}
\end{center}
\end{figure}                                         

Notice that as the distance between the sheets is increased the number of bands necessary to achieve the convergence in the screening $W$ and in the expansion of the Green Function $G$ rapidly increase  as in a simple quantum well where the number of bands below a give energy threshold increases with the system size. Therefore we maintained a fix cutoff in energy for all the calculations and increase the number of bands when necessary.\\
Another important parameter is the k-point grid. Here we report some convergences tests, see fig. \ref{conv_k}, for the gap correction at \boldmath$M$\unboldmath as function of the k-point sampling. We want underline two important aspects: first even if the convergence of the exchange part of the self-energy can be very slow fig. \ref{conv_k} panel $b)$, the combination of exchange and correlation part converges quite rapidly with the k-points; second for doped graphene, the convergence becomes harder with increasing doping because a larger Fermi surface has to be sampled.

\addcontentsline{toc}{chapter}{Bibliography}
\bibliographystyle{apsrev}
\bibliography{graphene.bib}

\begin{thebibliography}{25}
\expandafter\ifx\csname natexlab\endcsname\relax\def\natexlab#1{#1}\fi
\expandafter\ifx\csname bibnamefont\endcsname\relax
  \def\bibnamefont#1{#1}\fi
\expandafter\ifx\csname bibfnamefont\endcsname\relax
  \def\bibfnamefont#1{#1}\fi
\expandafter\ifx\csname citenamefont\endcsname\relax
  \def\citenamefont#1{#1}\fi
\expandafter\ifx\csname url\endcsname\relax
  \def\url#1{\texttt{#1}}\fi
\expandafter\ifx\csname urlprefix\endcsname\relax\def\urlprefix{URL }\fi
\providecommand{\bibinfo}[2]{#2}
\providecommand{\eprint}[2][]{\url{#2}}

\bibitem[{\citenamefont{Novoselov et~al.}(2005)\citenamefont{Novoselov, Geim,
  V., Jiang, Katsnelson, Grigorieva, Dubonos, and Firsov}}]{novoselov2}
\bibinfo{author}{\bibfnamefont{K.}~\bibnamefont{Novoselov}},
  \bibinfo{author}{\bibfnamefont{A.~K.} \bibnamefont{Geim}},
  \bibinfo{author}{\bibfnamefont{M.~S.} \bibnamefont{V.}},
  \bibinfo{author}{\bibfnamefont{D.}~\bibnamefont{Jiang}},
  \bibinfo{author}{\bibfnamefont{M.~I.} \bibnamefont{Katsnelson}},
  \bibinfo{author}{\bibfnamefont{I.~V.} \bibnamefont{Grigorieva}},
  \bibinfo{author}{\bibfnamefont{S.~V.} \bibnamefont{Dubonos}},
  \bibnamefont{and} \bibinfo{author}{\bibfnamefont{A.~A.}
  \bibnamefont{Firsov}}, \bibinfo{journal}{Nature}
  \textbf{\bibinfo{volume}{438}}, \bibinfo{pages}{197} (\bibinfo{year}{2005}).

\bibitem[{\citenamefont{Novoselov}(2004)}]{novoselov1}
\bibinfo{author}{\bibfnamefont{K.}~\bibnamefont{Novoselov}},
  \bibinfo{journal}{Science} \textbf{\bibinfo{volume}{306}},
  \bibinfo{pages}{666} (\bibinfo{year}{2004}).

\bibitem[{\citenamefont{Zhang et~al.}(2005)\citenamefont{Zhang, Tan, Stormer,
  and Kim}}]{zhang}
\bibinfo{author}{\bibfnamefont{Y.}~\bibnamefont{Zhang}},
  \bibinfo{author}{\bibfnamefont{Y.~W.} \bibnamefont{Tan}},
  \bibinfo{author}{\bibfnamefont{H.}~\bibnamefont{Stormer}}, \bibnamefont{and}
  \bibinfo{author}{\bibfnamefont{P.}~\bibnamefont{Kim}},
  \bibinfo{journal}{Nature} \textbf{\bibinfo{volume}{438}},
  \bibinfo{pages}{201} (\bibinfo{year}{2005}).

\bibitem[{\citenamefont{Westervelt}(2008)}]{nano}
\bibinfo{author}{\bibfnamefont{R.~M.} \bibnamefont{Westervelt}},
  \bibinfo{journal}{Science} \textbf{\bibinfo{volume}{320}},
  \bibinfo{pages}{5874} (\bibinfo{year}{2008}).

\bibitem[{\citenamefont{Bostwick et~al.}(2007)\citenamefont{Bostwick, Ohta,
  Seyller, Horn, and Rotenberg}}]{bostwick}
\bibinfo{author}{\bibfnamefont{A.}~\bibnamefont{Bostwick}},
  \bibinfo{author}{\bibfnamefont{T.}~\bibnamefont{Ohta}},
  \bibinfo{author}{\bibfnamefont{T.}~\bibnamefont{Seyller}},
  \bibinfo{author}{\bibfnamefont{K.}~\bibnamefont{Horn}}, \bibnamefont{and}
  \bibinfo{author}{\bibfnamefont{E.}~\bibnamefont{Rotenberg}},
  \bibinfo{journal}{Nat. Phys.} \textbf{\bibinfo{volume}{2}},
  \bibinfo{pages}{595} (\bibinfo{year}{2007}).

\bibitem[{\citenamefont{Gr\"uneis et~al.}(2008)\citenamefont{Gr\"uneis,
  Attaccalite, Pichler, Zabolotnyy, Shiozawa, Molodtsov, Inosov, Koitzsch,
  Knupfer, Schiessling et~al.}}]{alexander}
\bibinfo{author}{\bibfnamefont{A.}~\bibnamefont{Gr\"uneis}},
  \bibinfo{author}{\bibfnamefont{C.}~\bibnamefont{Attaccalite}},
  \bibinfo{author}{\bibfnamefont{T.}~\bibnamefont{Pichler}},
  \bibinfo{author}{\bibfnamefont{V.}~\bibnamefont{Zabolotnyy}},
  \bibinfo{author}{\bibfnamefont{H.}~\bibnamefont{Shiozawa}},
  \bibinfo{author}{\bibfnamefont{S.}~\bibnamefont{Molodtsov}},
  \bibinfo{author}{\bibfnamefont{D.}~\bibnamefont{Inosov}},
  \bibinfo{author}{\bibfnamefont{A.}~\bibnamefont{Koitzsch}},
  \bibinfo{author}{\bibfnamefont{M.}~\bibnamefont{Knupfer}},
  \bibinfo{author}{\bibfnamefont{J.}~\bibnamefont{Schiessling}},
  \bibnamefont{et~al.}, \bibinfo{journal}{Phys. Rev. Lett.}
  \textbf{\bibinfo{volume}{100}}, \bibinfo{pages}{037601}
  (\bibinfo{year}{2008}).

\bibitem[{\citenamefont{Gr\"uneis et~al.}(2007)\citenamefont{Gr\"uneis,
  Pichler, Shiozawa, Attaccalite, Wirtz, Molodtsov, Follath, Weber, and
  A.}}]{alexander2}
\bibinfo{author}{\bibfnamefont{A.}~\bibnamefont{Gr\"uneis}},
  \bibinfo{author}{\bibfnamefont{T.}~\bibnamefont{Pichler}},
  \bibinfo{author}{\bibfnamefont{H.}~\bibnamefont{Shiozawa}},
  \bibinfo{author}{\bibfnamefont{C.}~\bibnamefont{Attaccalite}},
  \bibinfo{author}{\bibfnamefont{L.}~\bibnamefont{Wirtz}},
  \bibinfo{author}{\bibfnamefont{S.}~\bibnamefont{Molodtsov}},
  \bibinfo{author}{\bibfnamefont{R.}~\bibnamefont{Follath}},
  \bibinfo{author}{\bibfnamefont{R.}~\bibnamefont{Weber}}, \bibnamefont{and}
  \bibinfo{author}{\bibfnamefont{R.}~\bibnamefont{A.}}, \bibinfo{journal}{Phys.
  Status Solidi(b)} \textbf{\bibinfo{volume}{244}}, \bibinfo{pages}{41294133}
  (\bibinfo{year}{2007}).

\bibitem[{\citenamefont{McChesney et~al.}(2007)\citenamefont{McChesney,
  Bostwick, Ohta, Emtsev, Seyller, Horn, and Rotenberg}}]{rotenberg}
\bibinfo{author}{\bibfnamefont{J.~L.} \bibnamefont{McChesney}},
  \bibinfo{author}{\bibfnamefont{A.}~\bibnamefont{Bostwick}},
  \bibinfo{author}{\bibfnamefont{T.}~\bibnamefont{Ohta}},
  \bibinfo{author}{\bibfnamefont{K.~V.} \bibnamefont{Emtsev}},
  \bibinfo{author}{\bibfnamefont{T.}~\bibnamefont{Seyller}},
  \bibinfo{author}{\bibfnamefont{K.}~\bibnamefont{Horn}}, \bibnamefont{and}
  \bibinfo{author}{\bibfnamefont{E.}~\bibnamefont{Rotenberg}}
  (\bibinfo{year}{2007}), \bibinfo{note}{arXiv:0705.3264v1}.

\bibitem[{\citenamefont{Pietronero and Strassler}(1981)}]{pietronero}
\bibinfo{author}{\bibfnamefont{L.}~\bibnamefont{Pietronero}} \bibnamefont{and}
  \bibinfo{author}{\bibfnamefont{S.}~\bibnamefont{Strassler}},
  \bibinfo{journal}{Phys. Rev. Lett.} \textbf{\bibinfo{volume}{8}},
  \bibinfo{pages}{593} (\bibinfo{year}{1981}).

\bibitem[{\citenamefont{Das~Sarma et~al.}(2007)\citenamefont{Das~Sarma, Hu,
  Hwang, and Tse}}]{sarma}
\bibinfo{author}{\bibfnamefont{S.}~\bibnamefont{Das~Sarma}},
  \bibinfo{author}{\bibfnamefont{B.~Y.} \bibnamefont{Hu}},
  \bibinfo{author}{\bibfnamefont{E.}~\bibnamefont{Hwang}}, \bibnamefont{and}
  \bibinfo{author}{\bibfnamefont{W.}~\bibnamefont{Tse}},
  \bibinfo{journal}{Phys. Rev. B.} \textbf{\bibinfo{volume}{75}},
  \bibinfo{pages}{121406} (\bibinfo{year}{2007}).

\bibitem[{\citenamefont{Polini et~al.}(2008)\citenamefont{Polini, Azgari,
  Borghi, Barlas, Pereg-Barnea, and MacDonald}}]{polini}
\bibinfo{author}{\bibfnamefont{M.}~\bibnamefont{Polini}},
  \bibinfo{author}{\bibfnamefont{R.}~\bibnamefont{Azgari}},
  \bibinfo{author}{\bibfnamefont{G.}~\bibnamefont{Borghi}},
  \bibinfo{author}{\bibfnamefont{Y.}~\bibnamefont{Barlas}},
  \bibinfo{author}{\bibfnamefont{T.}~\bibnamefont{Pereg-Barnea}},
  \bibnamefont{and}
  \bibinfo{author}{\bibfnamefont{A.}~\bibnamefont{MacDonald}},
  \bibinfo{journal}{Phys. Rev. B.} \textbf{\bibinfo{volume}{77}},
  \bibinfo{pages}{081411} (\bibinfo{year}{2008}).

\bibitem[{\citenamefont{Mishchenko}(2007)}]{mishchenko}
\bibinfo{author}{\bibfnamefont{E.~G.} \bibnamefont{Mishchenko}},
  \bibinfo{journal}{Phys. Rev. Lett.} \textbf{\bibinfo{volume}{98}},
  \bibinfo{pages}{216801} (\bibinfo{year}{2007}).

\bibitem[{\citenamefont{Gonze et~al.}(2002)\citenamefont{Gonze, Beuken,
  Caracas, Detraux, Fuchs, Rignanese, Sindic, Verstraete, Zerah, Jollet
  et~al.}}]{ABINIT}
\bibinfo{author}{\bibfnamefont{X.}~\bibnamefont{Gonze}},
  \bibinfo{author}{\bibfnamefont{J.-M.} \bibnamefont{Beuken}},
  \bibinfo{author}{\bibfnamefont{R.}~\bibnamefont{Caracas}},
  \bibinfo{author}{\bibfnamefont{F.}~\bibnamefont{Detraux}},
  \bibinfo{author}{\bibfnamefont{M.}~\bibnamefont{Fuchs}},
  \bibinfo{author}{\bibfnamefont{G.-M.} \bibnamefont{Rignanese}},
  \bibinfo{author}{\bibfnamefont{L.}~\bibnamefont{Sindic}},
  \bibinfo{author}{\bibfnamefont{M.}~\bibnamefont{Verstraete}},
  \bibinfo{author}{\bibfnamefont{G.}~\bibnamefont{Zerah}},
  \bibinfo{author}{\bibfnamefont{F.}~\bibnamefont{Jollet}},
  \bibnamefont{et~al.}, \bibinfo{journal}{Comp. Mat. Sci.}
  \textbf{\bibinfo{volume}{25}}, \bibinfo{pages}{478} (\bibinfo{year}{2002}).

\bibitem[{\citenamefont{Troullier and Martins}(1991)}]{pseudo}
\bibinfo{author}{\bibfnamefont{N.}~\bibnamefont{Troullier}} \bibnamefont{and}
  \bibinfo{author}{\bibfnamefont{J.~L.} \bibnamefont{Martins}},
  \bibinfo{journal}{Phys. Rev. B.} \textbf{\bibinfo{volume}{43}},
  \bibinfo{pages}{1993} (\bibinfo{year}{1991}).

\bibitem[{\citenamefont{Aryasetiawan and Gunnarsson}(1998)}]{gunnarsson}
\bibinfo{author}{\bibfnamefont{F.}~\bibnamefont{Aryasetiawan}}
  \bibnamefont{and}
  \bibinfo{author}{\bibfnamefont{O.}~\bibnamefont{Gunnarsson}},
  \bibinfo{journal}{Rep. Prog. Phys.} \textbf{\bibinfo{volume}{61}},
  \bibinfo{pages}{237} (\bibinfo{year}{1998}).

\bibitem[{\citenamefont{Onida et~al.}(2002)\citenamefont{Onida, Reining, and
  Rubio}}]{rubio}
\bibinfo{author}{\bibfnamefont{G.}~\bibnamefont{Onida}},
  \bibinfo{author}{\bibfnamefont{L.}~\bibnamefont{Reining}}, \bibnamefont{and}
  \bibinfo{author}{\bibfnamefont{A.}~\bibnamefont{Rubio}},
  \bibinfo{journal}{Rev. Mod. Phys.} \textbf{\bibinfo{volume}{74}},
  \bibinfo{pages}{601} (\bibinfo{year}{2002}).

\bibitem[{yam()}]{yambo}
\bibinfo{note}{A. Marini et al. YAMBO is a FORTRAN/C code for Many-Body
  calculations in solid state and molecular physics: http://yambo-code.org/}.

\bibitem[{gwf()}]{gwfull}
\bibinfo{note}{Note that different studies showed that performing a full GW
  without vertex correction can produce worse result that a single iteration,
  see for instance the discussion in \cite{gunnarsson}}.

\bibitem[{\citenamefont{Hybertsen and Louie}(1986)}]{ppmodel}
\bibinfo{author}{\bibfnamefont{M.}~\bibnamefont{Hybertsen}} \bibnamefont{and}
  \bibinfo{author}{\bibfnamefont{S.}~\bibnamefont{Louie}},
  \bibinfo{journal}{Phys. Rev. B.} \textbf{\bibinfo{volume}{34}},
  \bibinfo{pages}{5390} (\bibinfo{year}{1986}).

\bibitem[{\citenamefont{Trevisanutto et~al.}(2008)\citenamefont{Trevisanutto,
  Giorgetti, Reining, M., and Olevano}}]{olevano}
\bibinfo{author}{\bibfnamefont{P.~E.} \bibnamefont{Trevisanutto}},
  \bibinfo{author}{\bibfnamefont{C.}~\bibnamefont{Giorgetti}},
  \bibinfo{author}{\bibfnamefont{L.}~\bibnamefont{Reining}},
  \bibinfo{author}{\bibfnamefont{L.}~\bibnamefont{M.}}, \bibnamefont{and}
  \bibinfo{author}{\bibfnamefont{V.}~\bibnamefont{Olevano}}
  (\bibinfo{year}{2008}), \bibinfo{note}{arXiv:0806.3365v1}.

\bibitem[{\citenamefont{Zhou et~al.}(2008)\citenamefont{Zhou, Siegel, Fedorov,
  and Lanzara}}]{zhou}
\bibinfo{author}{\bibfnamefont{S.~Y.} \bibnamefont{Zhou}},
  \bibinfo{author}{\bibfnamefont{D.~A.} \bibnamefont{Siegel}},
  \bibinfo{author}{\bibfnamefont{A.}~\bibnamefont{Fedorov}}, \bibnamefont{and}
  \bibinfo{author}{\bibfnamefont{A.}~\bibnamefont{Lanzara}}
  (\bibinfo{year}{2008}), \bibinfo{note}{arXiv:0801.3862}.

\bibitem[{\citenamefont{Giovannetti et~al.}(2007)\citenamefont{Giovannetti,
  Khomyakov, Brocks, Kelly, and van~den Brink}}]{giovannetti}
\bibinfo{author}{\bibfnamefont{G.}~\bibnamefont{Giovannetti}},
  \bibinfo{author}{\bibfnamefont{P.}~\bibnamefont{Khomyakov}},
  \bibinfo{author}{\bibfnamefont{G.}~\bibnamefont{Brocks}},
  \bibinfo{author}{\bibfnamefont{P.}~\bibnamefont{Kelly}}, \bibnamefont{and}
  \bibinfo{author}{\bibfnamefont{J.}~\bibnamefont{van~den Brink}},
  \bibinfo{journal}{Phys. Rev. B.} \textbf{\bibinfo{volume}{76}},
  \bibinfo{pages}{0731303} (\bibinfo{year}{2007}).

\bibitem[{\citenamefont{Hwang and Das~Sarma}(2008)}]{hwang}
\bibinfo{author}{\bibfnamefont{E.}~\bibnamefont{Hwang}} \bibnamefont{and}
  \bibinfo{author}{\bibfnamefont{S.}~\bibnamefont{Das~Sarma}},
  \bibinfo{journal}{Phys. Rev. B.} \textbf{\bibinfo{volume}{77}},
  \bibinfo{pages}{081412} (\bibinfo{year}{2008}).

\bibitem[{\citenamefont{Roldan et~al.}(2008)\citenamefont{Roldan, Lopez-Sancho,
  and Guinea}}]{guinea}
\bibinfo{author}{\bibfnamefont{R.}~\bibnamefont{Roldan}},
  \bibinfo{author}{\bibfnamefont{M.}~\bibnamefont{Lopez-Sancho}},
  \bibnamefont{and} \bibinfo{author}{\bibfnamefont{F.}~\bibnamefont{Guinea}},
  \bibinfo{journal}{Phys. Rev. B} \textbf{\bibinfo{volume}{77}},
  \bibinfo{pages}{115410} (\bibinfo{year}{2008}).

\bibitem[{\citenamefont{Calandra and Mauri}(2007)}]{mauri}
\bibinfo{author}{\bibfnamefont{M.}~\bibnamefont{Calandra}} \bibnamefont{and}
  \bibinfo{author}{\bibfnamefont{F.}~\bibnamefont{Mauri}},
  \bibinfo{journal}{Phys. Rev. B} \textbf{\bibinfo{volume}{76}},
  \bibinfo{pages}{161406} (\bibinfo{year}{2007}).

\end{thebibliography}
\end{document}